\documentclass{iau_FM}
\usepackage{graphicx}
\usepackage{tabularx}

\title[FM9] %% give here short title %%
{Measures of luminous and dark matter\\ in galaxies across time}

\author[Freundlich, Sharma, Thater, et al.]   %% give here short author list %%
{
Jonathan Freundlich$^1$\thanks{Jonathan Freundlich, Gauri Sharma, and Sabine Thater were the chairs of the Focus Meeting, in alphabetical order. They are followed by the members of the Scientific Organisation Committee. The focus meeting was initiated by Gauri Sharma; the rationale was written in concert; the highlights by Jonathan Freundlich.}, 
Gauri Sharma$^1$$\dagger$,
Sabine Thater$^2$$\dagger$,\\
Mousumi Das$^3$,
Benoit Famaey$^1$,
Katherine Freese$^4$, 
Marie Korsaga$^{5, 6}$,
Julien Lavalle$^7$,
Chung Pei Ma$^8$,
Moses Mogotsi$^9$,
Cristina Popescu$^{10}$,
Francesca Rizzo$^{11}$,
Laura V. Sales$^{12}$, 
Miguel A. S\'anchez-Conde$^{13, 14}$,
Glenn van de Ven$^{2}$,
Hongsheng Zhao$^{15}$,
and Alice Zocchi$^{2}$
}

\affiliation{
\flushleft
$^1$    Observatoire astronomique, Universit\'e de Strasbourg, CNRS UMR 7550, Strasbourg, France\\
$^2$    Department of Astrophysics, University of Vienna, Vienna, Austria\\
$^3$    Indian Institute of Astrophysics, Bangalore, Karnataka, India\\
$^4$    Department of Physics, The University of Texas at Austin, USA\\
$^5$    Universit\'e Joseph Ki-Zerbo, Burkina Faso\\
$^6$    Instituto de Astrof\'isica de Andaluc\'ia (IAA-CSIC), Granada, Spain\\
$^7$    Laboratoire Univers et Particules de Montpellier, CNRS \& Universit\'e de Montpellier, France\\
$^8$    Departments of Astronomy and Physics, University of California, Berkeley, California, USA\\
$^9$    South African Astronomical Observatory (SAAO), Cape Town, South Africa\\
$^{10}$ University of Central Lancashire, Jeremiah Horrocks Institute, Preston, UK\\
$^{11}$ Kapteyn Astronomical Institute, University of Groningen, Groningen, Netherlands\\
$^{12}$    %Department of Physics and Astronomy, University of California, California, USA\\
Department of Physics and Astronomy, University of California, Riverside, CA, USA\\
$^{13}$ Departamento de F\'isica Te\'orica, M-15, Universidad Aut\'onoma de Madrid, Madrid, Spain\\
$^{14}$ Instituto de F\'isica Te\'orica UAM-CSIC, Universidad Aut\'onoma de Madrid, Madrid, Spain\\
$^{15}$ Scottish Universities Physics Alliance, University of Saint Andrews, Saint Andrews, Fife, UK\\
}

\pubyear{2024}
\jname{Astronomy in Focus, Focus Meeting 9} 
\editors{Diana M.~Worrall, ed.}
\begin{document}

\maketitle

\begin{abstract}
Dark matter is one of the pillars of the current standard model of structure formation: it is assumed to constitute most of the matter in the Universe. However, it can so far only be probed indirectly through its gravitational effects, and its nature remains elusive. In this focus meeting, we discussed different methods used to estimate galaxies' visible and dark matter masses in the nearby and distant Universe. We reviewed successes of the standard model relying on cold dark matter, confronted observations with  simulations, and highlighted inconsistencies between the two. We discussed how robust mass measurements can help plan, perform, and refine particle dark matter searches. We further exchanged about alternatives to cold dark matter, such as warm, self-interacting, and fuzzy dark matter, as well as modified gravity. Finally, we discussed prospects and strategies that could be implemented to reveal the nature of this crucial component of the Universe.

\keywords{dark matter, galaxies, galaxies: halos, galaxies: kinematics and dynamics}
\end{abstract}

\firstsection % if your document starts with a section,

%\newpage
\section{Rationale}

\medskip

In 1933, Fritz Zwicky used the Doppler velocities and luminosities of galaxies in the Coma cluster to estimate the total mass \cite[(Zwicky 1933)]{Zwicky1933}. He reported an enormous discrepancy between, respectively, the gravitating and luminous masses of the system. This study raised the question of a `missing mass' and coined the term `dark matter'. Although the discrepancy was overestimated, still a factor of six remains today. In the late 1970s and beginning 1980s, extended rotation curves showed that the rotation velocity of disc galaxies does not decrease beyond their visible domain (\cite[Rubin et al. 1970]{Rubin1970}, \cite[1980]{Rubin1980}, \cite[Bosma 1978]{Bosma1978}, \cite[van Albada et al. 1985]{vanAlbada1985}). That is, the enclosed gravitating mass increases as a function of radius, suggesting that galaxies are surrounded by a dark matter halo which extends much further than their visible matter. Theoretical models of structure formation \cite[(e.g., White \& Rees 1978,  Peebles 1982)]{White1978,Peebles1982}, which were gradually refined over the years thanks to observations of the Universe on large scales, have in parallel also revealed  the need for such a non-baryonic and cold matter component, making it one of the main pillars of the current cosmological model (e.g. \cite[Cirelli, Strumia \& Zupan 2024]{Cirelli2024}, and references therein).
\\

%Theoretical models of structure formation supported these observational findings \cite[(White \& Rees 1978)]{White1978}, and dark matter has since then become one of the main pillars of the current cosmological model.

The last 50 years saw the obtention of a large number of detailed observations of galaxies, clusters and the cosmic web across the electromagnetic spectrum. 
%In the last 50 years, astronomy has provided a large number of detailed observations of galaxies, clusters and the cosmic web across the electromagnetic spectrum. 
These observations enabled to measure velocities of gas and stars in galaxies to trace the gravitating mass, and luminosities to infer the visible mass, along with maps of the distribution and chemistry of gas and stars. %All these observations and the associated scaling relations between the properties of galaxies suggest the presence of dark matter, which would act via gravity. Cosmologically, it would need to represent 84\% of the matter content of the Universe, but at small scales its fraction varies greatly from system to system.
%fill 26\% of the Universe, whereas baryonic matter would only account for about 5\% \cite[(Planck collaboration 2018)]{Planck2018}. Dark matter would typically dominate the total mass of galaxies, but its nature remains unknown.
At the same time, the increase in computing resources and techniques has enabled astrophysicists and cosmologists to simulate the Universe from large to small scales, to construct detailed dynamical models of galaxies, and to compare both these simulations and models to observations in a statistically robust way (e.g. \cite[Wechsler \& Tinker 2018]{Wechsler2018}, \cite[Angulo \& Hahn 2022]{Angulo}). 
All this has led to the current consensus that, without dark matter, it would be impossible to explain a plethora of phenomena such as the amplitude of baryonic acoustic oscillations, the formation and evolution of cosmological structures, and the motion of stars and gas on galactic scales. However, the {\it a priori} most obvious particle candidates \cite[(e.g., Porter, Johnson \& Graham 2011)]{Porter2011} have not yet been detected by non-gravitational means, and there might be observational hints for departure from the cold and collisionless dark matter paradigm on subgalactic scales \cite[(e.g. Bullock \& Boylan-Kolchin 2017)]{Bullock2017}, opening the ground for alternative theories of gravity. In parallel, many exotic candidates have been proposed for the nature of dark matter, from ultra-light axions to massive compact objects like primordial black holes. 
\\

The current era of precision astrophysics and cosmology provides us with an enormous amount of quality data that was impossible to obtain in the past. For example, spectroscopic surveys with JWST and ALMA can resolve the hot and cold gas kinematics of galaxies up to the peak of cosmic dawn; upcoming gamma-ray facilities such as CTA, in combination with the Fermi-LAT in orbit, will put to the test various dark matter particle models; new generation radio telescopes like MeerKAT and SKA will allow the study of neutral hydrogen up to higher redshifts. Are we sufficiently prepared to interpret these data? Do we have the models to robustly measure and disentangle the distribution of luminous and dark matter, and the simulations to accurately mimic galaxy-scale baryonic processes and their impact on the total mass distribution? Is there something lacking in the overall efforts? How do we improve as a community, and what should be the main focus for the next five years? 
%If so, then what keeps hindering the searches for the dark matter particle? 
%Finally, theoretical efforts are also strongly needed to better understand the effects of baryonic physics on small scales, beyond better observations. 
\\

% Julien Lavalle : Maybe emphasize that, beyond better observations, theoretical efforts are also strongly needed to better understand the effects of baryonic physics on small scales, which will anyway be hindered by (i) intrinsic difficulties (star formation from first principles in complex environments), (ii) scaling issues (relevant processes occur at AU scales while effects of interest are at the pc scale and beyond), which currently make cosmological simulations limited in power (a star is at best a $10^4$~M$_\odot$ particle in a simulation).
 
At the IAU General Assembly 2024 in Cape Town, South Africa, the Focus Meeting `Measures of Luminous and Dark Matter in Galaxies Across Time' drew the attention on galaxy mass estimates and their importance to understand the interplay between baryonic and dark matter and help unravel the nature of dark matter. 
We notably aimed at contributing to answer the following questions: 
(1) How consistent are galaxy mass measurements based on different tracers?
(2) How robust are galaxy mass measurements against modelling assumptions and degeneracies?
(3) How accurate are state-of-the-art cosmological simulations reproducing current galaxy mass measurements across scales and cosmic time?
(4) Are dark matter halo properties the result of baryonic feedback, or are they due to the nature of dark matter itself?
(5) What are the missing observations to understand the nature of dark matter?
(6) What is the impact of having accurate mass measurements on current dark matter particle constraints, and how will this help us to strategically plan (and succeed with) future dark matter searches?
(7) Do we need alternative theories of dark matter? If yes, how capable are they in representing galaxy mass measurements across scales and cosmic time?
\\

%The chairs of the focus meeting were Jonathan Freundlich, Gauri Sharma and Sabine Thater. The Scientific Organizing Commitee further included Mousumi Das, Benoit Famaey, Marie Korsaga, Julien Lavalle, Chung Pei Ma, Moses Mogotsi, Francesca Rizzo, Laura Sales, Miguel A. S\'anchez-Conde, Glenn van de Ven, Hongsheng Zhao, and Alice Zocchi. 

%\section{Program}

%\medskip

%The focus meeting aimed at addressing the crucial question of the nature of dark matter, at highlighting the necessity of robust mass determinations of galaxies across cosmic time, and at discussing the comparison of observational and experimental results with cosmological simulations. We also opened the floor for experts in alternative theories of dark matter and dedicated a session to discussing the strategies for future facilities and the optimal use of archival data. 
%
%The proposed program contained six sessions, covering the following topics. \\

%To address these questions, we proposed to host six sessions, covering the following topics: 
To address these questions, the proposed program contained six sessions, covering the following topics: 
\begin{enumerate}
    \item[\textbf{(1)}] \textbf{Observed distribution of dark matter.}
In this session, we planned to discuss the robustness of galaxy masses estimated using various tracers in local and high-redshift Universe, such as stellar and gas kinematics from galaxies (spirals, ellipticals, dwarfs, dwarf spheroidal, satellites), galaxy clusters, gravitational lensing, globular clusters, planetary nebulae, tidal-tails. We wanted to delve into the dynamical mass modelling techniques of the different tracers, and to address the following questions: 
How consistent are galaxy mass measurements based on different tracers?
How robust are galaxy mass measurements against modelling assumptions and degeneracies?
What are the missing observations to understand the nature of dark matter?

    \item[\textbf{(2)}] \textbf{Successes of the cold dark matter paradigm.} In this session, we planned to validate and connect the mass measurements of various tracers with cosmological simulations within the standard model of cosmology. This session also aimed to address the interplay between baryons and dark matter, to discuss open questions related to galaxy mass assembly and dark matter halo evolution, and to highlight the successes of the cold dark matter paradigm. We notably planned to address the following questions:
    How accurate are cold dark matter cosmological simulations in reproducing the current galaxy mass measurements across scales and cosmic time?
    To what extent are dark matter halo properties at galactic scales driven by baryonic processes or the nature of dark matter itself?

    \item[\textbf{(3)}] \textbf{Challenges of the cold dark matter paradigm.} In this session, we planned to continue the previous session focussing on the regimes where the standard model of cosmology faces difficulties in describing observations. We notably planned to address the following questions: 
    What are the challenges of the cold dark matter paradigm at galactic and extra-galactic scales?
    Are these challenges due to technological limitations, or does the theory need to be refined or revised?

    \item[\textbf{(4)}] \textbf{Constraining dark matter particle candidates with galactic observations.} In this session, we wanted to show the role of robust mass measurements of galaxies in planning, performing and refining particle dark matter searches. We notably wanted to discuss the competence of existing and upcoming gamma-ray facilities (such as CTA and Fermi-LAT) and how they plan to test dark matter particle models. With this session, we planned to address the following questions:
    What is the impact of having accurate mass measurements on current dark matter particle constraints, and how will this help us to plan (and succeed with) future dark matter searches?
    What is the potential of current and future gamma-ray telescopes to test dark matter particle models with galactic observations? 

    \item[\textbf{(5)}] \textbf{Alternatives to cold dark matter particles.} Given the status of the mass discrepancy problem in galaxies, efforts of dark matter searches, and the limitation of the cold dark matter paradigm, we planned to discuss the scope and success of alternative theories to cold dark matter such as warm, self-interacting, and fuzzy dark matter, and modified theories of gravity. In particular, we planned to discuss the following questions:
    How accurate are \textit{other} dark matter cosmological simulations in reproducing the current galaxy mass measurements across scales and cosmic time? 
    Do we need an alternative to dark matter, such as modified gravity? If yes, can it justify the observables of structure formation and Big-bang nucleosynthesis at all scales? 

    \item[\textbf{(6)}] \textbf{Future dark matter studies.} In this session, we planned to discuss the strategies for upcoming facilities: the designs of instruments that are needed to disentangle the various mass components in the inner region of galaxies across the electromagnetic spectrum and cosmic time, and the sensitivity and design of future experiments needed for dark matter searches. Furthermore, we wanted to discuss the optimal use of archival data in predicting and/or revealing the nature of dark matter, including big data management and machine learning. We notably planned to address the following questions:
    Which kind of experiments are needed to understand the nature of dark matter?
    How can we use archival data to better understand the mass discrepancy problem and improve the current state of dynamical models? 
    Can machine learning help strengthen dark matter constraints? 

\end{enumerate}

%\input{table}

%Following this program, we invited the six following speakers: Sedona Price for session (1), Kyle Oman for (2), Federico Lelli for (3), Vivianna Gammaldi for (4), Giulia Despali for (5), and Françoise Combes for (6). We received 107 contributions, from which we devised a program that included 24 contributed talks, 12 lightning talks, and 51 posters. Given the submitted contributions, planned sessions (1) and (2) were merged in the final program into two sessions revolving around observational constraints on dark matter in different environments, its evolution with cosmic time, and kinematic modeling techniques. Table \ref{tab1} displays the program of the focus meeting. The six sessions were chaired by Gauri Sharma, Jonathan Freundlich, Sedona Price, Federico Lelli, Françoise Combes, and Caroline Foster, respectively. 

\section{Highlights}

\medskip 
\subsection{\raggedright Observational constraints on dark matter across cosmic time }
\medskip 

The first session started with observational constraints on dark matter, at different epochs and in different types of galaxies. %, from dwarfs to the most massive spiral galaxies. 

{\bf $\bullet$ Sedona Price} ({\it see proceedings}) reviewed recent endeavours to measure galaxy dynamical masses and their dark matter fraction out to the peak epoch of star formation and even beyond, up to $z\sim3$, owing to optical and near-infrared spectroscopy as well as sub-mm interferometers (e.g. \cite[Wuyts et al. 2016]{Wuyts2016}, \cite[Genzel et al. 2017]{Genzel2017}, \cite[2020]{Genzel2020}, \cite[Price et al. 2020]{Price2020}, \cite[2021]{Price2021}, \cite[Bouch\'e et al. 2022]{Bouche2022}, \cite[Nestor Shachar et al. 2023]{Nestor2023}). These studies point towards an increase of the dark matter fraction within the effective radius with cosmic time, lower values of this quantity in dense, massive star-forming galaxies, and possible evidence for cored dark matter haloes at the peak epoch of star formation. The presentation also emphasized some of the challenges to measure the dark matter fraction, namely the need for deep enough data, the galaxy-halo degeneracy, the parameterization choices of mass and kinematic profiles, the gas pressure support correction, and the impact of non-circular motions. 

{\bf $\bullet$ Lai Yee Lilian Lee} presented the ALMA CRISTAL survey resolving [CII] kinematics of typical main sequence galaxies at $4<z<6$, which notably finds that about half of the targetted galaxies are rotating disks (Lee et al. 2024b, in prep.). Their kinematical modelling with the 3D foward modeling tool DysmalPy (\cite[Price et al. 2021]{Price2021}, Lee et al. 2024a, in prep.), which accounts for beam-smearing, spectral broadening, and projection effects, yields high velocity dispersions $\sigma$ and low $V_{\rm rot}/\sigma$ consistent with Toomre-unstable disks, as well as low dark matter fractions at galactic scale.

{\bf $\bullet$ Caroline Forster} showed results from the VLT/MUSE MAGPI Survey at $z\sim 0.3$ indicating no dependence of galaxy morphology, stellar and gas rotation on their environment (Foster et al., in prep.). 

{\bf $\bullet$ Richard McDermid} presented resolved stellar kinematics of massive galaxies from the MAGPI survey at $z\sim 0.3$ analysed with the Schwarzschild orbit-based modelling technique to recover intrinsic mass structures and dark matter fraction, yielding a larger diversity of density slopes than more simple Jeans models (\cite[Derkenne et al. 2024]{Derkenne2024}). 

{\bf $\bullet$ Nathan Deg} presented kinematic scaling relations, including the stellar and baryonic Tully Fisher relations, from a WALLABY pilot survey with atomic gas measurements from the Australian SKA Pathfinder (ASKAP), which gave a first taste of upcoming studies with the full WALLABY survey and eventually SKA (\cite[Murugeshan et al. 2024]{Murugeshan2024}). 

%\subsection{\raggedright The case of dwarf galaxies}
%\medskip 

%\subsection{\raggedright Dynamical scaling relations}

\medskip 
\subsection{\raggedright Kinematical modeling shortcomings and tests with numerical simulations}
\medskip 

The second session and other talks continued to present observational constraints on dark matter, with an emphasis on kinematic models, their shortcomings, and tests with numerical simulations. 

{\bf $\bullet$  Kyle Oman} ({\it see proceedings}) started by reviewing the cusp-core problem, its currrent status, and the remaining challenge of the diversity of central dark matter densities at fixed stellar mass (\cite[Oman et al. 2015]{Oman2015}). Possible solutions to the cusp-core problem include redistributing dark matter via dark matter physics (as for example with self-interacting dark matter), redistributing dark matter via baryonic physics (e.g. through the effect of supernova outflows on the gravitational potential), and modifying the law of gravity (e.g. MOND). But errors and uncertainties in kinematic models and their interpretation could also play a crucial role. On one side, dynamics may not always trace the potential due to perturbations in gas and/or stars (e.g. \cite[Downing \& Oman 2023]{Downing2023}); on the other, recovering a rotation curve from limited observational information is challenging, so reliable uncertainty estimates are crucial (e.g. \cite[Sellwood et al. 2021]{Sellwood2021}). 

{\bf $\bullet$ Zahra Basti} presented tests of observational procedures to derive the dark matter distribution using the NIHAO zoom-in cosmological hydrodynamical simulations. She showed that while these procedures are able to recover the simulated dark matter mass profiles at large radii, there can be a mismatch as large as 50\% near the center. The mismatch is attributed to non-circular motions and the structural assumptions that were made, such as the bulge/disk decomposition. 

{\bf $\bullet$ Nikhil Arora} highlighted how feedback from active galactic nuclei affects the dark matter distribution in the NIHAO simulations, suppressing as much as 40\% of the central dark matter mass (\cite[Arora et al. 2024]{Arora2024}). 

{\bf $\bullet$ Nondh Panithanpaisal} showed that subhaloes beyond the virial radius in the FIRE cosmological simulations consist of two distinct populations separable in phase space, namely splashback subhaloes that have once fallen into the host on one side, and field subhaloes that are on their first infall on the other side. Since splashback subhaloes show signs of tidal disruption that can bias dynamical mass estimates, it is important to distinguish the two populations when modelling their dynamics. 

{\bf $\bullet$ Michael Fall}, in place of {\bf Aaron Ludlow}, showed how spurious dynamical heating, due to the limited number of particles in simulated galaxies compared to real ones, can affect the kinematics and structure of galaxies in both isolated and cosmological simulations (\cite[Ludlow et al. 2021]{Ludlow2021}, \cite[2023]{Ludlow2023}, \cite[Wilkinson et al. 2023]{Wilkinson2023}). They proposed a semi-empirical model to identify the condition to neglect this insidious numerical effect. 

{\bf $\bullet$ Suchira Sarkar} ({\it see proceedings}) reported a population of galaxies with a central high surface brightness stellar disk surrounded by an extended low surface brighness stellar disk in the IllustrisTNG50 simulation, hosted by $\sim 10^{12}~M_\odot$ dark matter halos and harbouring extended gas reservoirs. These simulated galaxies appear to be similar to several observed giant low surface brightness galaxies.

\medskip 
\subsection{\raggedright Challenges of the cold dark matter paradigm}
\medskip 
 
%\subsection{LCDM challenges and simulations}

The third session delved into some of the challenges of the cold dark matter paradigm, with an emphasis on scaling relations observed between baryonic matter and the gravitational field. 

{\bf $\bullet$ Federico Lelli} ({\it see proceedings}) reviewed puzzling regularities in the relative distributions of baryons and dark matter across the galaxy population revealed by mass modelling, encapsulated in a set of dynamical laws: (1) remarkably flat rotation curves at large radii (\cite[Mistele et al. 2024]{Mistele2024}), (2) the baryonic Tully-Fisher relation linking the total baryonic mass with the flat rotation velocity at large radii, (3) the central surface density relation linking the baryonic surface density with the dynamical mass surface density at small radii, and (4) the radial acceleration relation linking the baryonic gravitational field with the observed centripetal acceleration at each radius (\cite[Lelli et al. 2017]{Lelli2017}, \cite[Lelli 2022]{Lelli2022}, \cite[2024]{Lelli2024}, \cite[McGaugh 2020]{McGaugh2020}). These relations are particularly tight and involve a characteristic acceleration scale. They constitute {\it a priori} predictions of MOND modified gravity, which could thus be the basis for an alternative to particle dark matter or provide an effective dark matter scaling relation yet to be understood. 

{\bf $\bullet$  Ji Hoon Kim} ({\it see proceedings}) investigated the central surface density and the baryonic Tully-Fisher relations as a function of the specific angular momentum, the spin parameter, and environment.  

{\bf $\bullet$ Marie Korsaga} further reported the universality of the HI-to-halo mass ratio for isolated disk galaxies in the local Universe, inferred from high-quality rotation curves, revealing another surprising regularity in the relative distribution of baryons and dark matter (\cite[Korsaga et al. 2023]{Korsaga2023}).

{\bf $\bullet$ Enrico di Teodoro} showed results about the dynamics and scaling relations of the most massive spiral galaxies in the local Universe, using H$\alpha$ slit spectroscopy and HI interferometric data to derive rotation curves and estimate their baryonic and non-baryonic masses (\cite[Di Teodoro et al. 2021]{DiTeodoro2021}, \cite[2023]{DiTeodoro2023}). These galaxies display no breaks nor bends in the stellar and baryonic Tully-Fisher relations, and only a slight non-significant bend in the Fall relations between stellar and baryonic angular momentum and mass, seemingly indicating that baryons in such disk galaxies retain the specific angular momentum of the dark matter halo.

{\bf $\bullet$ Michael Fall} reported that the dark matter haloes of late-type galaxies infered from extended HI rotation curves are usually more massive than those of early-type galaxies, at fixed stellar mass, such that their stellar-to-halo mass relations are very different: rising monotonically for late-type galaxies while declining rapidly beyond a stellar mass $\sim 3\times 10^{10}~M_\odot$ for early-type galaxies (\cite[Posti et al. 2019]{Posti2019}, \cite[Posti \& Fall 2021]{Posti2021}, \cite[Di Teodoro et al. 2023]{DiTeodoro2023}). 

{\bf $\bullet$ Pavel Mancera Pi\~na} presented a sample of gas-rich ultra-diffuse galaxies which are outliers of the baryonic Tully-Fisher relation, suggesting atypical dark matter distributions (\cite[Mancera Pi\~na et al. 2019]{ManceraPina2019}, \cite[2020]{ManceraPina2020}, \cite[2022]{ManceraPina2022}, \cite[2024]{ManceraPina2024}). High-resolution HI atomic gas observations of three such galaxies and their kinematical modeling yield particularly low-density and low-concentration halos that may be at odds with cold dark matter cosmological simulations but would notably be a natural outcome in the self-interacting dark matter paradigm. MOND and fuzzy dark matter were also tested, but disfavoured. This nevertheless shows the potential for testing the nature of dark matter with kinematical modeling of ultra-diffuse galaxies. 

\medskip 
\subsection{\raggedright Dark matter particle searches}
\medskip 

The fourth session explored dark matter particle searches and means to constrain dark matter particle candidates. 

{\bf $\bullet$ Donggeun Tak} reviewed the principle of indirect dark matter searches, based on the possibility that the annihilation and/or decay of dark matter would produce particles observable through their gamma-ray emission in high dark matter density regions, and indicated how nearby dwarf spheroidal galaxies may be one of the best targets to search such signatures given their high dark matter density and their lack of other gamma-ray sources. He then presented the ground-based Very Energetic Radiation Imaging Telescope Array System (VERITAS), constituted of four imagining atmospheric Cherenkov telescopes. Searches for weakly-interacting massive particle (WIMP) or ultra-heavy dark matter (UHDM) annihilation signals have so far been negative (\cite[Acharyya et al. 2024]{Acharrya2024}). 

{\bf $\bullet$ Carlo Nipoti} ({\it see proceedings}) reported how nearby dwarf spheroidal galaxies, such as the Milky Way satellites Fornax, Sculptor and Leo I, have sufficiently accurate estimates of their central dark matter density to estimate their so-called $J$- and $D$-factors, which are key  to interpret measurements of $\gamma$-ray flux in terms of particle physics properties of the dark matter annihilation and decay (\cite[Pascale et al. 2018]{Pascale2018}, \cite[2024]{Pascale2024} and in prep., Arroyo-Polonio et al. in prep.).

{\bf $\bullet$ Viviana Gammaldi} presented indirect dark matter searches in dwarf irregular galaxies using satellite and ground based observatories, such as Fermi-LAT, HESS and HAWC (e.g. \cite[Gammaldi et al. 2021]{Gammaldi2021}). Dwarf irregulars are indeed expected to display a high dark matter content and a low gamma-ray astrophysical background at relatively close distances. The upcoming Cherenkov Telescope Array should improve current constraints.

{\bf $\bullet$ Michael Sarkis} presented a software package that is able to quickly calculate emission profiles from a wide range of WIMP phenomenological models and astrophysical targets (\cite[Sarkis \& Beck 2024]{Sarkis2024}). It was applied to search for WIMP emissions in a sample of galaxy clusters observed with MeerKAT (\cite[Lavis et al. 2023]{Lavis2023}).

\medskip 
\subsection{\raggedright Alternatives to cold dark matter}
\medskip 

The fifth and sixth sessions explored alternatives to the cold dark matter paradigm, such as warm, self-interacting and fuzzy dark matter, modified gravity, and primordial black holes. 

{\bf $\bullet$ Giulia Despali} reviewed the main signatures of warm and self-interacting dark matter on galaxy formation and evolution. Warm dark matter notably yields fewer satellites and low-mass haloes, larger amounts of matter in filaments and walls and a delayed structure formation, which in practice affects the matter power-spectrum, satellite counts, gravitational lensing, and the Lyman-$\alpha$ forest. Self-interacting dark matter yields variations in the density profile of dark matter haloes, with either a central core or a very cuspy profile due to core collapse with high or velocity-dependent cross sections, a diversity in the properties of dwarfs and rotation curves, differences in galaxy dynamics and bars, rounder halo shapes, and specific effects on cluster mergers. Ongoing efforts to carry out cosmological simulations in these alternative frameworks were presented (e.g. \cite[Despali et al. 2020]{Despali2020}, \cite[2022]{Despali2022}, and in prep., \cite[Correa et al. 2024]{Correa2024}, Gutke et al. in prep.). 

{\bf $\bullet$ Françoise Combes} reviewed the two alternatives to cold dark matter that are MOND modified gravity and fuzzy dark matter. MOND provides good fits of rotation curves and naturally yields the Tully-Fisher relation and the radial acceleration relation. The Aether Scalar Tensor theory introduced by \cite[Skordis \&  Z\l{}o\'snik (2021)]{Skordis2021} could be a promising avenue to ground MOND in a more fundamental theory, notably as it reproduces the angular power spectrum of the Cosmic Microwave Background. MOND theories can be constrained by wide binaries and Solar System dynamics, but they face challenges in galaxy clusters (e.g. \cite[Freundlich et al. 2022]{Freundlich2022}, \cite[Nagesh et al. 2024]{Nagesh2024}). Fuzzy dark matter made of axion-like particles prevents structure formation below a certain mass due to quantum pressure, changes the aspect of the cosmic web with quantum interferences at large scales, and flattens dark matter density profiles at the center of dwarf halos, as shown by simulations (e.g. \cite[May \& Springel 2023]{May2023}, \cite[Nori et al. 2023]{Nori2023}). It also yields density fluctuations  that can heat up stellar structures and hence constrain the axion mass (e.g. \cite[Marsh \& Niemeyer 2019]{Marsh2019}, \cite[El-Zant et al. 2020]{El-Zant2020}). 

{\bf $\bullet$ Victor Robles} studied the effect of supernova feedback at the center of fuzzy dark matter halos using numerical simulations, showing that it leads to central densities that continuously fluctuate in time, even after feedback has ceased. The resulting time-dependent scatter translates to an uncertainty in the determination of the dark matter distribution (\cite[Robles et al. 2024]{Robles2024}).

{\bf $\bullet$ Przemek Mroz} presented the results of the 20 year search for long-timescale gravitational microlensing events in the Large Magellanic Cloud by the OGLE survey, which would have provided evidence for dark matter being composed of dark compact objects such as primordial black holes. No events with timescales longer than a year were found, which strongly constrain the fraction of dark matter that would be composed of compact objects (\cite[Mroz et al. 2024]{Mroz2024}). Thirteen microlensing events were detected with timescales shorter than a year, but they can be explained by stellar objects in the Large Magellanic Cloud or in the Milky Way.

\medskip 
\subsection{\raggedright Gravitational lensing}
\medskip

Talks in different sessions presented observational constraints on the dark matter distribution through strong and weak gravitational lensing and discussed lens modeling techniques. 

{\bf $\bullet$  John McKean} presented how high resolution imaging of gravitational lenses through Very Long Baseline Interferometry (VLBI) observations enables to test different dark matter models, including cold, warm and fuzzy dark matter, and presented evidence for perturbations due to dark matter subhalos (\cite[Powell et al. 2022]{Powell2022}, \cite[2023]{Powell2023}, \cite[Stacey et al. 2024]{Stacey2024}). The upcoming next generation Very Large Array (ngVLA) and SKA should enable further such tests of dark matter models (McKean et al. in prep). 

{\bf $\bullet$  Catherine Cerny} presented how the combination of stellar velocity dispersion measurements and strong lensing mass models can untangle the baryonic and dark matter distributions inside brightest cluster galaxies, providing evidence for dark matter cores in such galaxies but also highlighting the different modeling assumptions (Cerny et al. 2024, submitted).  

{\bf $\bullet$ Surhud More} presented work linking the stellar mass of galaxies and their dark matter halos using photometry and weak lensing observations from the Subaru Hyper Suprime Cam survey, finding that the galaxy-dark matter connection does not vary significantly over cosmic time (\cite[Chaurasiya et al. 2023]{Chaurasiya2023}). 

{\bf $\bullet$ Samuel Lange} presented the discovery of a dark matter substructure candidate from strong lens modelling with the JWST (Lange et al., in prep). 

{\bf $\bullet$ Nency Patel} presented the mass distribution of a complex galaxy cluster at $z=0.3$, Abel 2744, with both weak and strong gravitational lensing.  

{\bf $\bullet$ Carlos Melo-Carneiro} explored how the self-consistent modelling of stellar dynamics and lensing can improve measurements of total mass, stellar mass and dark matter content of early-type galaxies using mock observations generated from the IllustrisTNG50 high-resolution cosmological hydrodynamical simulations. He reported that joint models outperform the individual methods, but that they still struggle to fully recover dark matter and halo parameters (\cite[Melo-Carneiro et al. 2024]{Melo-Carneiro2024}). 

{\bf $\bullet$ Surajit Kalita} ({\it see proceedings}) explored how the gravitational lensing of fast radio bursts could be utilized to constrain the fraction of primordial black holes and modified gravity (\cite[Kalita et al. 2023]{Kalita2023}).

%%% Displaced from session 4

\medskip 
\subsection{\raggedright Probing dark matter with the stellar distribution}
\medskip 

Different talks proposed indirect means to probe dark matter and disentangle models through the stellar distribution, in particular stellar haloes, bars, tidal dwarf galaxies, or globular cluster populations. 

{\bf $\bullet$ Azadeh Fattahi} discussed how galactic stellar halos can help explore and test the predictions of different dark matter models (\cite[Forouhar-Moreno  et al. 2024]{Forouhar-Moreno2024}). Stellar halos are indeed formed from the accretion and disruption of dwarf galaxies, whose abundance, structure, and formation time differ from one dark matter model to another. Cosmological hydrodynamical simulations of Local-Group volumes run with cold, warm and self-interacting dark matter show that stellar halos in warm dark matter are very similar to those in cold dark matter, because their properties are set by only a few massive accretions despite the overall change in the abundance of dwarf galaxies, while stellar halos in self-interacting dark matter (with a $10~\rm cm^2/gr$ cross-section) have flatter density profiles and less radially biased orbits in the center. This is interpreted as due to the combination of shallower gravitational potentials for hosts and accreted objects, notably making the latter more susceptible to stripping, enhanced mass loss because of the self-interaction between host and satellite dark matter, and less efficient radialisation. 

{\bf $\bullet$ Francesca Fragkoudi} showed how bar dynamics can be used to constrain galaxy formation and dark matter models. Massive dark matter halos are indeed expected to stabilise the disc against bar formation and to slow down bars through dynamical friction  (e.g. \cite[White \& Rees 1978]{White1978}, \cite[Tremaine \& Weinberg 1984]{Tremaine1984}). Barred galaxies in the Auriga cold dark matter cosmological simulations have higher stellar masses than unbarred galaxies for a given halo mass, which leads to an offset in the Tully-Fisher relation (\cite[Fragkoudi et al. 2024]{Fragkoudi2024}). Detecting such an offset would provide evidence for dark matter and rule out modified gravity. The observation of fast bars is however not necessarily incompatible with the cold dark matter paradigm, provided that barred spiral galaxies are baryon-dominated, as is the case in the Auriga simulations but not in simulations with too-strong feedback (\cite[Fragkoudi et al. 2021]{Fragkoudi2021}). 

{\bf $\bullet$  Florent Renaud} showed using high resolution hydrodynamical simulations of interacting galaxies how galaxy mergers and tidal interactions can help probe dark matter and its alternatives. In particular, cold dark matter and MOND leave distinct imprints on observables like the starburst activity and the formation of tidal dwarf galaxies, with stronger and more spatially extended starbursts as well as more tidal dwarf galaxies in MOND (\cite[Renaud et al. 2014]{Renaud2014}, \cite[2015]{Renaud2015}, \cite[2016]{Renaud2016}). 

{\bf $\bullet$  Teymoor Saifollahi} indicated how globular clusters number counts could provide a secondary mass estimator for dwarf galaxies in the local Universe, given that classical dynamical mass estimators (stellar, globular cluster and gas dynamics) have limitations in the low-mass and dwarf regime, and presented how globular clusters are identified in data from the Euclid Space Telescope (\cite[Saifollahi et al. 2024]{Saifollahi2024}). 

{\bf $\bullet$ Lu\'isa Buzzo} addressed the formation and dark matter content of ultra-diffuse galaxies using both spectroscopy and their globular cluster population, hinting at different classes of ultra-diffuse galaxies: puffed-up dwarfs with regular globular cluster numbers and luminosity function, failed galaxies with larger globular cluster numbers, and possibly tidal or bullet-like dwarfs with irregular globular cluster luminosity function and dark matter content (\cite[Buzzo et al. 2024]{Buzzo2024} and in prep.).

{\bf $\bullet$  Jessica Doppel} discussed how well globular cluster abundance and kinematics enable to trace dark matter in dwarf galaxies using a catalog of globular clusters added in post-processing to the IllustrisTNG50 simulation (\cite[Doppel et al. 2024]{Doppel2024}). They notably find that globular clusters can in principle be a good tracer of the dark matter content, but that interloping intra-cluster globular clusters can cause overly-massive dark matter halos to be inferred in the case of ultra-diffuse galaxies. 

\medskip 
\subsection{\raggedright Perspectives with machine learning}
\medskip 

Different talks presented machine learning techniques to assess dark matter masses, which are likely to gain importance with the large amount of data which will be produced by upcoming facilities. 

{\bf $\bullet$ Connor Bottrell} presented a simulation-based inference model trained on a large sample of mock Subaru Hyper Suprime Cam images of galaxies formed in cosmological simulations to estimate stellar and halo masses as well their uncertainties (\cite[Bottrell et al. 2024]{Bottrell2024}, \cite[Hahn, Bottrell \& Lee 2024]{Hahn2024}). Using the luminosity, color and morphological information enables to significantly reduce the scatter in the stellar-to-halo mass relation. 

{\bf $\bullet$  Jiani Chu} investigated the ability of a multi-branch convolutional neural network to predict the central stellar and total masses as well as the stellar mass-to-light ratio from simulated galaxy multi-band images and stellar velocity maps from the IllustrisTNG100 simulations. Luminosity and velocity dispersion contribute most in predicting mass, star formation or color helps to predict the mass-to-light ratio, while stellar spin helps to predict the dark matter fraction, suggesting that current observations should be sufficient to decode the dynamical properties of galaxies (\cite[Chu et al. 2023]{Chu2023}). 

{\bf $\bullet$  Connor Stone} presented a GPU-accelerated tool to model gravitational lenses which does Bayesian inference using machine learning (\cite[Stone et al. 2024]{Stone2024}). 

{\bf $\bullet$  Anupreeta More} presented machine learning algorithms and a citizen science project that enabled to discover hundreds of strong gravitational lenses up to $z\sim 0.8$ in a Hyper Suprime Cam survey (\cite[More et al. 2024]{More2024}, \cite[Gawade et al. 2024]{Gawade2024}).

%\bigskip

\section{Conclusion}

\medskip

The focus meeting `Measures of luminous and dark matter in galaxies across time' at the IAU General Assembly 2024 in Cape Town covered different aspects pertaining to dark matter in the Universe. We discussed observational constraints across cosmic time and in different environments from kinematic modeling and gravitational lensing, shortcomings of the different methods, as well as tests and comparisons with numerical simulations. We delved into some of the challenges of the cold dark matter paradigm, dark matter particle searches, and alternatives to cold dark matter such as warm, self-interacting and fuzzy dark matter, modified gravity, and primordial black holes. Interestingly, different talks proposed indirect means to probe dark matter and disentangle models through the stellar distribution, namely through stellar haloes, bars, tidal dwarf galaxies, or globular cluster populations. Finally, a number of talks presented machine learning techniques to assess dark matter masses, which are poised to gain importance with upcoming facilities such as the Extremely Large Telescope, the next-generation Very Large Array, the Square Kilometer Array, and the Cherenkov Telescope Array. \\

\bigskip

%\subsection{Constraining dark matter particle candidates}

%\subsection{Alternatives to LCDM}

%\subsection{Perspectives and new techniques}

\noindent {\it Acknowledgements.}
J.F. would like to thank Françoise Combes and Federico Lelli for valuable comments on the manuscript. 

\end{document}